\documentclass[iop]{emulateapj}

\begin{document}

\title{Post-starburst Tidal Tails in the Archetypical Ultra Luminous Infrared
Galaxy Arp 220\altaffilmark{1}}

\author{Y. Taniguchi\altaffilmark{2},
K. Matsubayashi\altaffilmark{2},
M. Kajisawa\altaffilmark{2, 3},
Y. Shioya\altaffilmark{2},
Y. Ohyama\altaffilmark{4},
T. Nagao\altaffilmark{5, 6},
Y. Ideue\altaffilmark{2, 3}, 
T. Murayama\altaffilmark{7}, and
J. Koda\altaffilmark{8}}
\altaffiltext{1}{Based on data collected at the
         Subaru Telescope, which is operated by
         the National Astronomical Observatory of Japan.}
\altaffiltext{2}{Research Center for Space and Cosmic Evolution, Ehime
University, Bunkyo-cho 2-5, Matsuyama, Ehime 790-8577, Japan}
\altaffiltext{3}{Graduate School of Science and Engineering, Ehime University,
Bunkyo-cho 2-5, Matsuyama, Ehime 790-8577, Japan}
\altaffiltext{4}{Institute of Astronomy and Astrophysics, Academia Sinica, PO
Box 23-141, Taipei 10617, Taiwan, R.O.C}
\altaffiltext{5}{Department of Astronomy, Kyoto University, Kitashirakawa
Oiwakecho, Sakyo-ku, Kyoto 606-8502, Japan}
\altaffiltext{6}{The Hakubi Center for Advanced Research, Kyoto University,
Yoshida-Ushinomiya-cho, Sakyo-ku, Kyoto 606-8302, Japan}
\altaffiltext{7}{Astronomical Institute, Tohoku University, Aramaki, Aoba,
Sendai 980-8578, Japan}
\altaffiltext{8}{Department of Physics and Astronomy, SUNY Stony Brook, Stony
Brook, NY 11794-3800, USA}

\begin{abstract}

We present our new deep optical imaging and long-slit spectroscopy
for Arp 220 that is the archetypical ULIRG in the local universe.
Our sensitive H$\alpha$ imaging has newly revealed large-scale, H$\alpha$
absorption, i.e., post-starburst regions in this merger; one is found in the
eastern superbubble and the other is in the two tidal tails
that are clearly reveled in our deep optical imaging.
The size of H$\alpha$ absorption region in the eastern bubble is 
5 kpc $\times$ 7.5 kpc and the observed
H$\alpha$ equivalent widths are $\sim$ 2 \AA\, $\pm$ 0.2 \AA.
The sizes of the northern and southern H$\alpha$-absorption tidal tails
are $\sim$ 5 kpc $\times$ 10 kpc and $\sim$ 6 kpc $\times$ 20 kpc, respectively.
The observed H$\alpha$ equivalent widths range from 4 \AA\, to 7 \AA.
In order to explain the presence of the two post-starburst tails,
we suggest a possible multiple-merger scenario for Arp 220
in which two post-starburst
disk-like structures merged into one, and then caused the two tails.
This favors that Arp 220 is a multiple merging system composed of four 
or more galaxies, arising from a compact group of galaxies.
Taking our new results into account, we discuss a star formation history 
in the last 1 Gyr in Arp 220.

\end{abstract}

\keywords{galaxies: individual (Arp 220) --- galaxies: interactions ---
galaxies: starburst}

\section{INTRODUCTION}

Ultra luminous infrared galaxies (ULIRGs) whose infrared (8--1000 $\mu$m) luminosity exceeds 10$^{12}$ $L_{\odot}$ are the sites where extraordinarily
active star formation occurs in their central regions, triggered by galaxy
mergers \citep{Sanders:1988}.
Since such galaxies have been also found as massive, dusty galaxies at high redshift,
providing substantial star formation in early universe \citep{LeFloch:2005}, it
is important to understand their physical origins unambiguously.

Originally, \citet{Sanders:1988} proposed that ULIRGs come from
a merger between two (gas-rich) galaxies.
On one hand, multiple merger scenarios have also been proposed
for the formation of ULIRGs based on their morphological and 
dynamical properties \citep{Taniguchi:1998a, Borne:2000, Haan:2011}.
Unlike well-defined, on-going merging systems such as the Antennae (NGC 4038 + NGC 4039),
most of ULIRGs appear to have complicated morphological properties of
 so-called advanced mergers (e.g., \citealt{Sanders:1996}).

Namely, the following questions have not yet been
unsettled unambiguously for the origin of ULIRGs. 
(1) How many galaxies were merged into one?
(2) Which types of galaxies were merged into one?
(3) How were their orbital parameters?
Since the merging of galaxies can smear out morphological properties 
of original galaxies that take part in the merger, it is generally difficult
to give firm answers to the above questions 
(e.g., \citealt{Weil:1996, Bournald:2007}). 

It is worthwhile noting that a major merger between two galaxies may be a site
of multiple mergers because any galaxies have their satellite galaxies
(\citealt{Wang:2012} and references therein).
For example, let us imagine a case of the merger between our Galaxy and M 31.
Since both galaxies have
two significant satellite galaxies (LMC and SMC for our Galaxy 
and M32 and NGC 205 for M31), this major merger includes six galaxies.
Since a minor merger also gives dynamical disturbance to its host
disk galaxy (e.g., \citealt{Mihos:1996, Taniguchi:1996a, Bournald:2007,
Villabos:2008}), it seems to have a unambiguous discrimination 
between a major merger between two galaxies and a multiple merger.
It is also worthwhile noting that numerical simulations of multiple mergers 
can make an Arp 220-like morphology during the course of mergers among 
six galaxies \citep{Weil:1996}.
A one-sided long tidal tail structure (e.g., Mrk 273)
can also be generated by a multiple merger model \citep{Bekki:2001}.

Among the ULIRGs in the local universe, Arp 220 is the archetypical advanced merger
\citep{Sanders:1988, Joseph:1985}. Arp 220 shows two starburst events
at least; one is the central super star clusters (i.e., the on-going starburst;
\citealt{Shaya:1994, Taniguchi:1998b}), and the other is the figure ``8" shaped
structure of ionized gas driven by a superwind \citep{Heckman:1987, Heckman:1990,
Heckman:1996}. It is also suggested that there is a hidden starburst core
in the central region of Arp 220 that is obscured seriously, $A_V \sim 30$ -- 50
mag (e.g., \citealt{Anantharamaiah:2000, Shioya:2001}). 
Indeed, multi-phase star formation events in Arp 220 and other ULIRGs 
have been identified in more recent imaging and spectroscopic investigations 
(e.g., \citealt{Wilson:2006, Rodriguez-Zaurin:2008, Rodriguez-Zaurin:2009, Soto:2010}).
Accordingly, Arp 220 appears to have experienced a long history of episodic 
starbursts during the course of merging within last $\sim$ 1 Gyr. 

This complicated star formation history in Arp 220 may come from 
a multiple mergers including more than two galaxies (e.g,
\citealt{Taniguchi:1998a, Borne:2000}).
It is generally difficult to explore the history of 
such multiple mergers in an advanced merger phase.
However, since Arp 220 shows the evident on-going and
past starburst events noted above together with a faint pair of tidal tails,
it is worthwhile investigating its star formation history in more details.
For this purpose, we have carried out deep H$\alpha$ imaging using the 
Subaru Telescope and a long-slit spectroscopic observation with the Keck II
Telescope.
Based on our new observations, we discuss a possible 
scenario for the star formation history in Arp 220 and re-visit a multiple
merger scenario.

We use the distance of 77.6 Mpc toward Arp 220 that is estimated with both
the systemic velocity of 5434 km s$^{-1}$
\citep{deVaucouleurs:1991} and a 
Hubble constant $H_0$ = 70 km s$^{-1}$ Mpc$^{-1}$.
At this distance, 1\arcsec\, corresponds to 376 pc.

\section{OBSERVATIONS}

\subsection{Optical Imaging}

In order to understand the whole star formation history in Arp 220, one
efficient method is to reveal post-starburst regions (their characteristic
ages are $\sim$ 0.2 -- 1 Gyr) as well as the on-going active star forming
regions (their characteristic ages are $<$ 10 Myr).
For this purpose, we obtained deep H$\alpha$ on-band and off-band frames using
the Faint Object Camera And Spectrograph (FOCAS) on the Subaru Telescope on
2002 September 15 (UT).
Total integration time for the on and off bands were 8400 seconds and 2400
seconds, respectively.
The center wavelengths and bandwidths were 6709 \AA\, and 85 \AA\, 
and 6588 \AA\, and 73 \AA\, for the on and off bands, respectively.
The on-band frames include stellar continuum emission and
H$\alpha$+[\ion{N}{2}]$\lambda\lambda$6548,6583 line emission, while the
off-band frames include stellar continuum only.

Bias subtraction, flat-fielding, and sky subtraction were performed.
Flux calibration was carried out by using SDSS spectrum \citep{Abazajian:2009}
of a galaxy in the same field of view.
In continuum subtraction, we took consideration into the continuum slope
between the H$\alpha$ on and off wavelengths estimated from Keck spectrum
around eastern H$\alpha$ absorption region (see Section \ref{sec:result}).
This correction was applied to the whole region of the off-band image, and
reduced continuum intensity by 2 \% or H$\alpha$ equivalent width by $\sim$ 1.5
\AA\, than if we assume no continuum slope correction, or flat continuum.
Since the eastern H$\alpha$ absorption region and the two tidal tails have the same
$r' - i'$ color indices (Section \ref{sec:result}), this continuum correction
is reasonable.
The spatial resolution of FOCAS images was 0\arcsec.7.

Deep $R$-band frames of Arp 220 were also obtained by Subaru Prime Focus Camera
(Suprime-Cam) on 2004 February 18 (UT).
Bias subtraction, flat-fielding, and sky subtraction were performed.
Total integration time was 4200 seconds, and the spatial resolution was
1\arcsec.0.
This $R$-band image was used to find faint tidal features around Arp 220 (see
also, \citealt{Koda:2009}).

\subsection{Optical Spectroscopy}

A new optical spectrum was obtained by the Low Resolution Imaging Spectrometer
\citep[LRIS: ][]{Oke:1995} with the Keck II Telescope on 1997 April 16 (UT).
A 900-second exposure was taken with the 600 line mm$^{-1}$ grating, with a
central wavelength setting of 6024 \AA.
The grating is blazed at 5000 \AA\, and the dispersion is 1.28 \AA\,
pixel$^{-1}$.
The observed wavelength range and the spectral resolution were 4670 -- 7210 \AA\,
and 6.9 \AA, respectively.
The 1\arcsec\, slit with a length of 66\arcsec\, was oriented at the position
angle (PA) of 130 degree to observe both the nucleus of Arp 220 and its visual
companion D (seen at SE of Arp 220) simultaneously.
The spectrum was reduced by using IRAF with the standard manner, i.e.,
wavelength calibration, flat-fielding, and flux calibration (see
\citealt{Ohyama:1999} for detail).

\section{RESULTS}
\label{sec:result}

\subsection{Imaging Data}

In post starburst regions, massive OB stars already ceased and thus
intermediate-mass stars such as A-type stars are dominant populations.
Such regions are probed by strong hydrogen Balmer absorption lines (e.g.,
\citealt{Taniguchi:1996a}).

\begin{figure}
\epsscale{.90}
\plotone{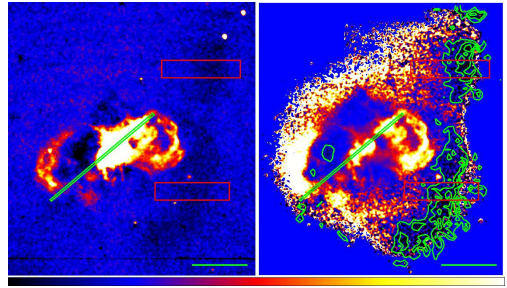}
\caption{(left panel) Continuum-subtracted, smoothed
H$\alpha$+[\ion{N}{2}]$\lambda\lambda$6548,6583 intensity map, and (right
panel) H$\alpha$ equivalent width map of Arp 220 obtained by FOCAS.
The color ranges are from $-1$ $\times$ 10$^{-18}$ to 5 $\times$ 10$^{-18}$ erg
cm$^{-2}$ s$^{-1}$ (0.20 arcsec)$^{-2}$ for H$\alpha$+[\ion{N}{2}] intensity
map, and from 10 to $-30$ \AA\, (positive value means absorption) for H$\alpha$
equivalent width map.
The contours are shown at 2, 6, and 10 \AA\, in the H$\alpha$ equivalent width
in absorption.
North is up and east is left.
The slit position in the longslit spectroscopic observation is shown with the
diagonal green box.
Two red rectangles represent the positions where continuum surface brightness
and equivalent widths are shown in Figure \ref{fig:filament-profile}.
Green bar in each panel at lower right corresponds to 10 kpc.
\label{fig:image-focas}}
\end{figure}

Figure \ref{fig:image-focas} shows the continuum-subtracted, smoothed
H$\alpha$+[\ion{N}{2}]$\lambda\lambda$6548,6583 image, and the H$\alpha$
equivalent width maps obtained by FOCAS, respectively.
We detect strong H$\alpha$ emission from the central region of Arp 220 and
shell-like structures in both SE and NW directions, as already reported in the
literature \citep{Heckman:1987, Heckman:1990, Heckman:1996}.
The strong H$\alpha$ emission from the central region is considered to be
attributed to the on-going starburst activity there.
On the other hand, the shell-like structures is due to the 
recent-past superwind, which is
a galactic-scale outflow caused by many supernovae.

\begin{figure}
\epsscale{.90}
\plotone{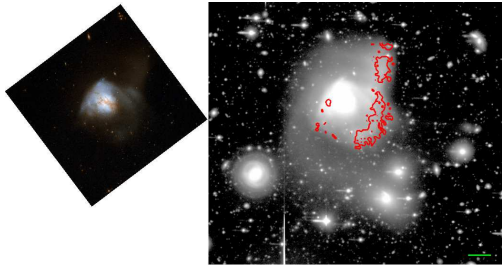}
\caption{$R$-band image of Arp 220 obtained by Suprime-Cam (right panel) with
the post starburst regions shown by red contours.
For comparison, the HST/ACS image of Arp 220 is also shown in the left panel
(http://hubblesite.org/gallery/album/pr2008016aq/).
North is up and east is left.
The post starburst regions are shown by red-color contours.
Green bar corresponds to 10 kpc.
\label{fig:image-scam}}
\end{figure}

\begin{figure}
\epsscale{.90}
\plotone{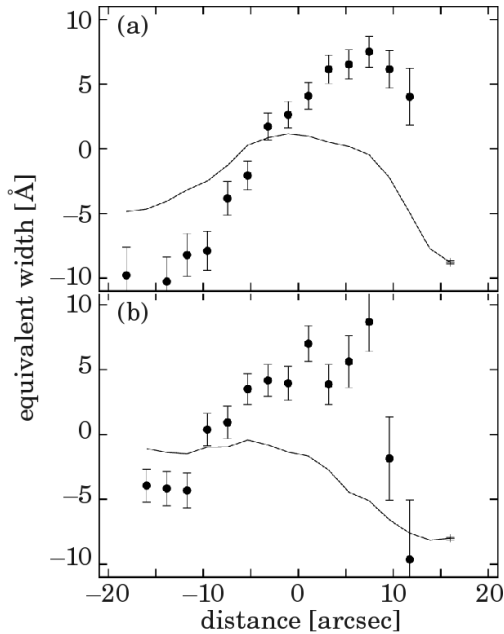}
\caption{Continuum surface brightness and H$\alpha$ equivalent width profiles
at (a) northern and (b) southern tails using FOCAS data.
Solid lines represent continuum surface brightness in arbitrary unit ($-10$
\AA\, corresponds to no flux), while filled circles represent equivalent
widths.
The silt positions are shown in Figure \ref{fig:image-focas}.
The aperture size in each data point is 2\arcsec\, $\times$ 8\arcsec.
Uncertainties in continuum surface brightness are almost the same at all
positions, and only the rightmost one is shown.
\label{fig:filament-profile}}
\end{figure}

On closer inspection, we find three H$\alpha$ absorption regions.
One H$\alpha$ absorption region is detected about 20\arcsec\, east from the
center, inside the eastern H$\alpha$ bubble.
The observed equivalent width of H$\alpha$ in the center 3\arcsec\, $\times$
3\arcsec\, is 2.3 \AA\, $\pm$ 0.15 \AA\, (1$\sigma$).
The size of this region is approximately 5 kpc $\times$ 7.5 kpc.
The other two spectacular H$\alpha$ absorption regions appear to be associated
with the two optically-faint tidal tails found in our deep $R$-band image (see
also Figure \ref{fig:image-scam}).
The observed equivalent width of H$\alpha$ amounts to $\sim$ 5 $\pm$ 0.4 \AA\,
in total; $\sim$ 4 $\pm$ 1.1 \AA\, at the continuum flux peak, and $\sim$ 7
$\pm$ 1.7 \AA\, at the outer side of the tails (Figure
\ref{fig:filament-profile}).
The sizes of the northern and southern H$\alpha$ absorption regions are $\sim$
5 kpc $\times$ 10 kpc and $\sim$ 6 kpc $\times$ 20 kpc, respectively.
Since these three regions show the H$\alpha$ absorption feature, A-type stars,
or post-starburst populations, are dominant in the optical continuum without
H$\alpha$ emission sources such as OB stars. 

\begin{figure}
\epsscale{.90}
\plotone{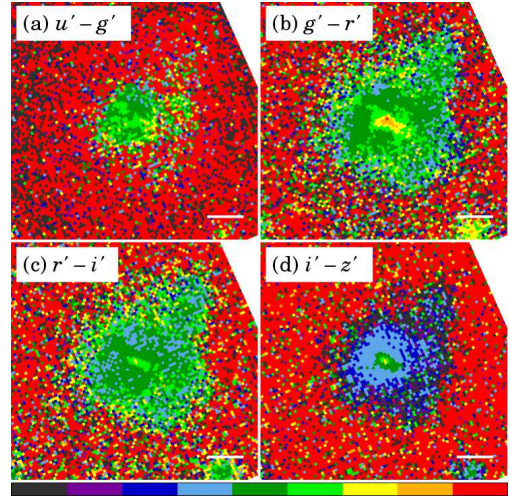}
\caption{(a) $u' - g'$, (b) $g' - r'$, (c) $r' - i'$, and (d) $i' - z'$ color
index maps obtained from SDSS images.
The color scales are from $-1$ to 3 mag for $u' - g'$ map, while from $-1$ to 1
mag for the other maps.
North is up and east is left.
White bar in each panel at lower right corresponds to 10 kpc.
\label{fig:image-sdss}}
\end{figure}

Figure \ref{fig:image-sdss} displays Arp 220 color index maps obtained from
SDSS images \citep{Abazajian:2009}.
We find that the three H$\alpha$ absorption regions have almost the same color
indices.
Furthermore, their color indices within several kpc except for the nuclear 
region are similar to those of the H$\alpha$ absorption regions.
These facts suggest that the dominant stellar population 
at these regions are the same,
i.e., post-starburst, which is consistent with \citet{Rodriguez-Zaurin:2008}
and \citet{Soto:2010} results.

Finally, we comment on an alternative interpretation that H$\alpha$ absorption
in the two tidal tails could be caused by the high extinction found in a number
of ULIRGs (e.g., \citealt{Rodriguez-Zaurin:2009}).
If the extinction is very high enough to hide on-going massive star formation,
the observed optical light might be dominated by intermediate-type stars,
resulting in strong H$\alpha$ absorption with little emission line.
Although such situations may occur in the central parts of ULIRGs including Arp
220 \citep{Shioya:2001}, it seems unlikely that such very dusty starbursts
occurred in the entire regions of both the eastern bubble and the two tidal
tails.
Therefore, we consider that the post-starburst interpretation is a more robust
idea for the observed H$\alpha$ absorption regions in Arp 220.

\subsection{Spectroscopic Data}
\label{sec:result-spec}

The so-called Lick index \citep{Worthey:1997} provides a common method for
calculating an equivalent width.
We therefore calculated the H$\beta$ Lick index as a function of
slit position.
H$\alpha$ equivalent width was also calculated in the same way as that of
Lick index.
Note that we adopt the following wavelength coverage; 6552.4 -- 6564.6 \AA\,
for the on-H$\alpha$ region, 6505.9 -- 6525.44 \AA\, and 6603.7 -- 6664.9 \AA\,
for the off-H$\alpha$ region.
We chose these ranges in order to avoid the effects from
[\ion{N}{2}]$\lambda\lambda$6548,6583 emission from the object, other 
absorption lines, and sky emission.

\begin{figure}
\epsscale{.90}
\plotone{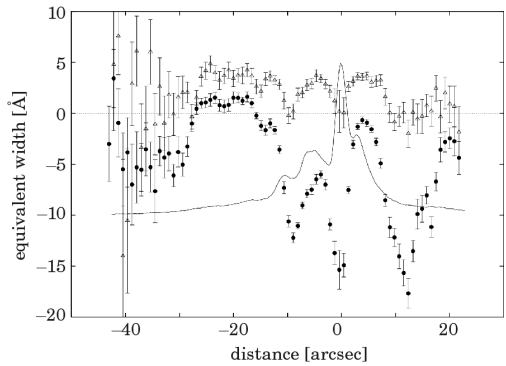}
\caption{Continuum flux density and equivalent width profiles along the slit
displayed in Figure \ref{fig:image-focas}, using LRIS spectroscopic data.
The horizontal axis represents the distance from the continuum peak in optical
wavelength.
Positive value in horizontal axis means in the northwest of the nucleus.
Solid line represents the surface brightness of continuum in arbitrary units
($-10$ \AA\, corresponds to no flux).
Filled circles and open triangles represent the H$\alpha$ equivalent widths and
H$\beta$ Lick index, respectively.
Positive equivalent widths mean absorption, while negative ones emission.
\label{fig:ew-profile}}
\end{figure}

Figure \ref{fig:ew-profile} represents continuum flux density and equivalent
width profiles along the slit position displayed in Figure
\ref{fig:image-focas}.
The H$\alpha$ equivalent width at the southern part of the eastern bubble is
$\sim$ 1.5 \AA, i.e., H$\alpha$ absorption, which is consistent with FOCAS
equivalent width map (Figure \ref{fig:filament-profile}).
The H$\beta$ equivalent width is more sensitive to post-starburst population
than that of H$\alpha$, because the flux of the H$\beta$ emission line is
generally about 3 times smaller than the H$\alpha$ flux (e.g.,
\citealt{Osterbrock:2006}).
The observed H$\beta$ Lick index of the eastern H$\alpha$ absorption region is
3 -- 4 \AA.
H$\beta$ absorption is detected not only around eastern bubble but also in the
central region of Arp 220, $\sim$ $+5$\arcsec\, and $\sim$ $-5$\arcsec\, from
the center (Figure \ref{fig:ew-profile}).
This fact suggests that post-starburst populations spread to several kpc
in diameter,
which is consistent with the results of both \citet{Rodriguez-Zaurin:2008} 
and \citet{Soto:2010}.
Unfortunately, we cannot measure the H$\beta$ equivalent width of the west
H$\alpha$ absorption tails, because they were out of the field of view of LRIS.

\section{DISCUSSION}

\subsection{A Multiple Merger Scenario}

Based on its morphology, it is obvious that 
Arp 220 is now in an advanced phase of a galaxy merger
\citep{Sanders:1988}. 
For Arp 220, \citet{Taniguchi:1998a} have discussed a possibility of 
multiple mergers.
Their argument is based on these facts that Arp 220 has two compact starburst 
regions and that a pair of nuclei (i.e., a pair of supermassive black
holes) is necessary to initiate the so-called 
nuclear starburst in a merging system in either a minor or a major merger
\citep{Taniguchi:1996b}.
This may be a plausible idea because 
the starburst is intrinsically different from 
ordinary star formation in galactic disks \citep{Daddi:2010}. 
The presence of multiple (at least three) OH megamaser sources can be
interpreted as evidence for major mergers of four galaxies in Arp 220 
\citep{Diamond:1989}. However, although further observations detected 
multiple components, it is reminded that they do not necessarily 
suggest the multiple nuclei \citep{Rovilos:2003}.

One interesting advantage of the multiple merger model by 
\citet{Taniguchi:1998a} is that they predicted the presence of 
a pair of counter rotating dense molecular gas disks 
before the actual observational detection by \citet{Sakamoto:1999}.
Arp 220 is a very gas rich system with $\sim 10^{10} M_{\odot}$
and a half of gas is associated with two apparent eastern and
western nuclei \citep{Scoville:1997}. 
It seems difficult to estimate orbital parameters of merging 
galaxies to explain all these observational results.

However, the post-starburst tidal tails revealed by our
our new observations presented here 
suggest a possible new idea on the merging history in Arp 220.
For further consideration of multiple mergers, 
we discuss the origin of Arp 220.

Here we focus on the origin of two post-starburst tidal tails.
First, in antenna-like tidal tails, one tidal tail emanates from one colliding
galaxy and thus two well-developed tidal tails are formed \citep{Toomre:1972}.
Taking this into account, we can postulate that two post-starburst systems
merged into one and then the observed two post-starburst tails were formed 
in Arp 220.
The observed smooth distribution of H$\alpha$ absorption features requires this
scenario although small-scale star formation activity could occur in gaseous
tidal tails \citep{Barnes:1992, Duc:1998}.

It should be noted that the size of each post-starburst system must be as
large as several kpc at least, corresponding to a typical scale of disk
galaxies.
Let us consider a probable star formation history in a major merger
between two disk galaxies.
During the course of such merger evolution, enhanced star formation events occurred,
being triggered by inward gas flow driven by non-axisymmetries in the galaxy disks
(e.g., \citealt{diMatteo:2007}). 
However, it should be reminded that the starburst events are essentially
confined in the nuclear region of each galaxy; i.e., 
within only a few kpc diameter area. Note that
this is comparable to those of nuclear starburst regions in disk galaxies
(e.g., \citealt{Balzano:1983}).
In order to make a pair of very large-scale ($\sim$ 10 kpc),
post-starburst tidal tails observed in Arp 220,
we need two large-scale ($\sim$ several kpc) post-starburst disk-like structures.
Such a structure can be made in an advanced phase of a merger between two
disk galaxies; i.e., a major merger remnant. 
We therefore strongly suggest that the observed two tidal tails 
in Arp 220 need a merger between two advanced (i.e., post-starburst) merger remnants.
Namely, we need four disk galaxies to explain the observed post-starburst
tidal tails in Arp 220.
We then conclude that Arp 220 comes not from a
typical merger between two galaxies but from a multiple merger including four
galaxies at least.

\begin{figure}
\epsscale{.90}
\plotone{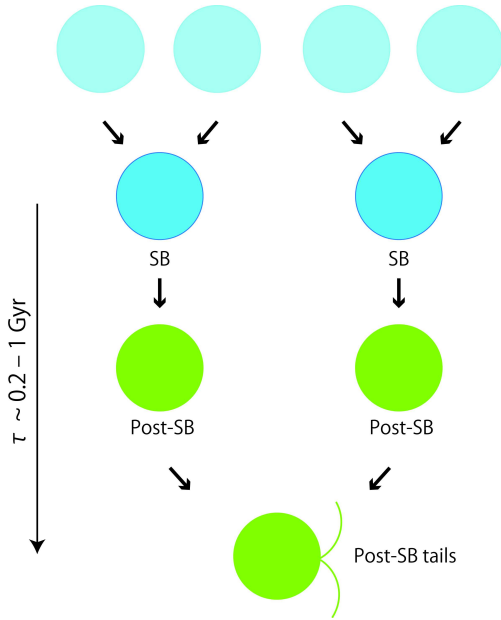}
\caption{A possible scenario for the formation of the post-starburst tails.
(1) Two merging partners caused an intense starburst.
Here we postulate that a couple of intense starbursts (SB) occurred in the
merging system about 0.2 -- 1 Gyr ago.
(2) After the cease of the starburst, these regions evolved into the
post-starburst (post-SB) regions.
And, (3) they merged into one and then the two post-SB tails were formed. 
\label{fig:arp220-scenario}}
\end{figure}

We estimate the travelling time of stars in the tidal tails from the
merger center to the present place of tails.
Given both the velocity of tidally-liberated stars as 100 km s$^{-1}$ and the
distance as 20 kpc, the travelling time is calculated as 200 Myr, being
consistent with the lifetime of post-starburst phase (0.2 -- 1 Gyr).
Our scenario for the evolution of Arp 220 is summarized in Figure 
\ref{fig:arp220-scenario}.

\subsection{Star Formation History in Arp 220}

Finally, we consider a possible unified picture for various star formation
events in Arp 220.
As shown in Figure 1, there are two more important starburst features in Arp
220; one is the central starburst and the other is the pair of superbubble
caused by the superwind activity \citep{Heckman:1987, Heckman:1990,
Heckman:1996, Colina:2004}.
The central starburst must be occurred very recently, $<$ 10 Myr ago, being the
life time of massive OB stars taken into account.
On the other hand, since the dynamical timescale of the observed superbubble is
estimated as a few tens Myr (e.g., \citealt{Heckman:1990}), Arp 220 experienced
another starburst a few tens Myr ago. These star formation events correspond
to those found in \citet{Rodriguez-Zaurin:2008} and \citet{Soto:2010}; i.e.,
the young stellar population in \citet{Rodriguez-Zaurin:2008} 
and the constant-star-formation region occurring in the central kpc region of 
Arp 220 \citep{Soto:2010}.

\begin{figure}
\epsscale{.90}
\plotone{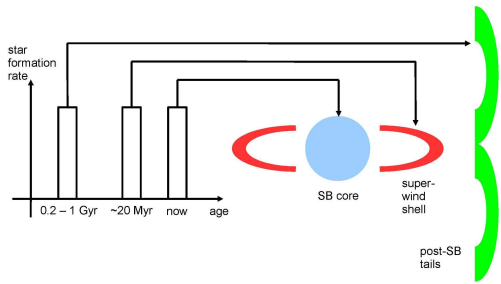}
\caption{A possible unified picture for various star formation activities in
Arp 220.
\label{fig:arp220-picture}}
\end{figure}

In summary, taking account of the two post-starburst tails, we propose a
unified evolution model for Arp 220 as shown in Figure \ref{fig:arp220-picture}.
These three starburst events could be driven by sequential multiple mergers in
Arp 220.

\section{CONCLUDING REMARKS}

There are a significant number of compact groups of
galaxies in the local universe \citep{Hickson:1982a, Hickson:1982b}, 
leading to multiple mergers.
Such compact groups of galaxies are also present in rich groups and 
outskirt regions of cluster of galaxies at intermediate redshift 
\citep{Mendel:2011}.
Moreover, a possible multiple merging system has been also found at 
$z >$ 1 \citep{Schawinski:2011}.

ULIRGs are considered to evolve to quasars and then giant early type galaxies
(elliptical or S0 galaxies).
A large number of high-redshift analogs of ULIRGs, so-called submm galaxies
(SMGs) have been discovered beyond $z \sim$ 1 (e.g., \citealt{Hughes:1998,
Barger:1998, Amblard:2011, Magdis:2011}).
Although major mergers between only two galaxies have been often 
considered in terms of the hierarchical evolution of galaxies from 
high redshift to the present day, we will have to take the impact of
 multiple mergers into account in future consideration.

\acknowledgments

We would like to thank the staff of Subaru Telescope for their professional
help in our observations.
We would also like to thank the HST/COSMOS team members because the deep
$R$-band image of Arp 220 was taken during a break of our COSMOS observing runs
on the Subaru Telescope; in particular, Nick Scoville, Dave Sanders, Bahram
Mobasher, Herve Aussel, and Peter Capak.
We would like to thank Dave Sanders and Kirsten Larson for useful discussion
on the global optical color properties of Arp 220.
Finally, we would like to thank the referees for his/her useful comments and
suggestions that improved this paper.
This work was financially supported in part by the Japan Society for the
Promotion of Science (Nos. 17253001, 19340046, 23244031, and 23654068).


\begin{thebibliography}{}

\bibitem[Abazajian et al.(2009)]{Abazajian:2009} Abazajian, K.~N.,
  Adelman-McCarthy, J.~K., Ag{\"u}eros, M.~A., et al.\ 2009, \apjs, 
  182, 543
\bibitem[Amblard et al.(2011)]{Amblard:2011} Amblard, A., Cooray, A., Serra,
  P., et al.\ 2011, \nat, 470, 510
\bibitem[Anantharamaiah et al.(2000)]{Anantharamaiah:2000} 
  Anantharamaiah, K.
  R., Viallefond, F., Mohan, N. R., Goss, W. M., \& Zhao, J. H. 
 \ 2000, \apj, 537, 613
\bibitem[Balzano(1983)]{Balzano:1983} Balzano, V.~A.\ 1983, \apj, 268, 602
\bibitem[Barnes \& Hernquist(1992)]{Barnes:1992} Barnes, J.~E., \& 
   Hernquist, L.\ 1992, \nat, 360, 715
\bibitem[Barger et al.(1998)]{Barger:1998} Barger, A.~J., Cowie, L.~L.,
  Sanders, D.~B., et al.\ 1998, \nat, 394, 248
\bibitem[Bekki(2001)]{Bekki:2001} Bekki, K., 2001, \apj, 564, 189
\bibitem[Borne et al.(2000)]{Borne:2000} Borne, K.~D., Bushouse, H.,
  Lucas, R.~A., \& Colina, L.\ 2000, \apjl, 529, L77
\bibitem[Bournald et al.(2007)]{Bournald:2007}Bournald, F.,
  Jog, C., \& Combes, F. 2007, \aap, 475, 1179
\bibitem[Colina et al.(2004)]{Colina:2004} Colina, L., Arribas, S., \& 
  Clements, D. \ 2004, \apj, 602, 181
\bibitem[Daddi et al.(2010)]{Daddi:2010}Daddi, E., et al. 2010, ApJ, 
  714, L118
\bibitem[Diamond et al.(1989)]{Diamond:1989}Diamond, P. J., Norris, R. P.,
  Baan, W. A., \& Booth, R. S. 1989, ApJ, 340, l49
\bibitem[de Vaucouleurs et al.(1991)]{deVaucouleurs:1991} 
  de Vaucouleurs, G.,de Vaucouleurs, A., Corwin, H., et al. 1991, 
  Third Reference Catalogue of
  Bright Galaxies (New York: Springer)
\bibitem[Di Matteo et al.(2007)]{diMatteo:2007} Di Matteo, P., 
  Combes, F., Melchior, A. -L., \& Semelin, B.\ 2007, \aap, 468, 61
\bibitem[Duc \& Mirabel(1998)]{Duc:1998} Duc, P.-A., \& 
  Mirabel, I.~F.\ 1998, \aap, 333, 813
\bibitem[Haan et al.(2011)]{Haan:2011} Haan, S., Surace, J.~A., Armus, L.,
  et al.\ 2011, \aj, 141, 100
\bibitem[Heckman et al.(1987)]{Heckman:1987} Heckman, T.~M., Armus, L.,
  \& Miley, G.~K.\ 1987, \aj, 93, 276
\bibitem[Heckman et al.(1990)]{Heckman:1990} Heckman, T.~M., Armus, L., 
  \& Miley, G.~K.\ 1990, \apjs, 74, 833 
\bibitem[Heckman et al.(1996)]{Heckman:1996} Heckman, T.~M., Dahlem, M., 
  Eales, S.~A., Fabbiano, G., \& Weaver, K.\ 1996, \apj, 457, 616 
\bibitem[Hickson(1982a)]{Hickson:1982a} Hickson, P.\ 1982a, \apj, 
   255, 382
\bibitem[Hickson(1982b)]{Hickson:1982b} Hickson, P.\ 1982b, \apj, 
   259, 930 
\bibitem[Hughes et al.(1998)]{Hughes:1998} Hughes, D.~H., Serjeant, S., 
   Dunlop, J., et al.\ 1998, \nat, 394, 241
\bibitem[Joseph \& Wright(1985)]{Joseph:1985} Joseph, R.~D., \& 
  Wright, G.~S.\ 1985, \mnras, 214, 87
\bibitem[Koda et al.(2009)]{Koda:2009} Koda, J., Scoville, N., 
   Taniguchi, Y., 
  \& Subaru COSMOS team 2009, American Astronomical Society Meeting Abstracts
  \#214, 214, \#418.03
\bibitem[Le Floc'h et al.(2005)]{LeFloch:2005} Le Floc'h, E., Papovich, 
  C., Dole, H., et al.\ 2005, \apj, 632, 169
\bibitem[Magdis et al.(2011)]{Magdis:2011} Magdis, G.~E., Elbaz, D., 
  Hwang, H.~S., Pep Team, \& Hermes Team 2011, Galaxy Evolution: 
  Infrared to Millimeter Wavelength Perspective, 446, 221
\bibitem[Mendel et al.(2011)]{Mendel:2011} Mendel, J.~T., Ellison, S.~L.,
  Simard, L., Patton, D.~R., \& McConnachie, A.~W.\ 2011, \mnras, 418, 1409
\bibitem[Mihos \& Hernquist(1996)]{Mihos:1996} Mihos, J.~C., \& Hernquist, L.\
  1996, \apj, 464, 641
\bibitem[Ohyama et al.(1999)]{Ohyama:1999} Ohyama, Y., Taniguchi, Y., 
   Hibbard, J.~E., \& Vacca, W.~D.\ 1999, \aj, 117, 2617
\bibitem[Oke et al.(1995)]{Oke:1995} Oke, J.~B., Cohen, J.~G., Carr, M.,
  et al.\ 1995, \pasp, 107, 375
\bibitem[Osterbrock \& Ferland(2006)]{Osterbrock:2006} Osterbrock,
   D.~E., \& Ferland, G.~J.\ 2006, Astrophysics of gaseous nebulae 
   and active galactic nuclei, 2nd.~ed.~by D.E.~Osterbrock a
   nd G.J.~Ferland.~Sausalito, CA:
  University Science Books, 2006,
\bibitem[Rodr{\'{\i}}guez Zaur{\'{\i}}n et al.(2008)]{Rodriguez-Zaurin:2008} 
  Rodr{\'{\i}}guez Zaur{\'{\i}}n, J., Tadhunter, C.~N., \& Gonz{\'a}lez
  Delgado, R.~M.\ 2008, \mnras, 384, 875
\bibitem[Rodr{\'{\i}}guez Zaur{\'{\i}}n et al.(2009)]{Rodriguez-Zaurin:2009} 
  Rodr{\'{\i}}guez Zaur{\'{\i}}n, J., Tadhunter, C.~N., \& Gonz{\'a}lez
  Delgado, R.~M.\ 2009, \mnras, 400, 1139
\bibitem[Rovilos et al.(2003)]{Rovilos:2003}Rovilos, E.,
    Diamond, P. J., Lonsdale, C. J., Lonsdale, C. J., \& Smith, H. E.
    2003, MNRAS, 342, 373
\bibitem[Sakamoto et al.(1999)]{Sakamoto:1999}Sakamoto, K.,
   Scoville, N. Z., Yun, M. S., Crosas, M., Genzel, R., \&
   Tacconi, L. J. 1999, \apj, 514, 68 
\bibitem[Sanders et al.(1988)]{Sanders:1988} Sanders, D.~B., Soifer, B.~T.,
  Elias, J.~H., et al.\ 1988, \apj, 325, 74
\bibitem[Sanders \& Mirabel(1996)]{Sanders:1996} Sanders, D.~B., \& Mirable,
  I. F., 1996, ARA \& A\, 34, 749
\bibitem[Schawinski et al.(2011)]{Schawinski:2011} Schawinski, K., Urry, 
  M., Treister, E., et al.\ 2011, \apjl, 743, L37
\bibitem[Scoville et al.(1997)]{Scoville:1997}Scoville, N. Z.,
   Yun, M. S., \& Vryant, P. M. 1997, ApJ, 484, 702
\bibitem[Shaya et al.(1994)]{Shaya:1994} Shaya, E.~J., Dowling, 
  D.~M., Currie, D.~G., Faber, S.~M., \& Groth, E.~J.\ 1994, \aj, 107, 1675
\bibitem[Shioya, Trentham, \& Taniguchi(2001)]{Shioya:2001} Shioya, Y., 
  Trentham, N,, \& Taniguchi, Y. \ 2001, \apjl, 548, L29
\bibitem[Soto \& Martin(2010)]{Soto:2010} Soto, K. T., \& Martin, C. L. 
   2010, \apj, 716, 332
\bibitem[Taniguchi et al.(1996)]{Taniguchi:1996a} Taniguchi, Y., 
  Ohyama, Y., Yamada, T., Mouri, H., \& Yoshida, M.\ 1996, \apj, 467, 215
\bibitem[Taniguchi \& Wada(1996)]{Taniguchi:1996b} Taniguchi, Y., \& Wada, K.\
  1996, \apj, 469, 581
\bibitem[Taniguchi \& Shioya(1998)]{Taniguchi:1998a} Taniguchi, Y., \& 
  Shioya, Y.\ 1998, \apjl, 501, L167
\bibitem[Taniguchi, Trentahm \& Shioya(1998)]{Taniguchi:1998b} Taniguchi, Y.,
  Trentahma, N., \& Shioya, Y.\ 1998, \apjl, 504, L79
\bibitem[Toomre \& Toomre(1972)]{Toomre:1972} Toomre, A., \& 
  Toomre, J.\ 1972, \apj, 178, 623
\bibitem[Villabos \& Helmi(2008)]{Villabos:2008}Villabos, A., \&
  Helmi, A. 2008, MNRAS, 391, 1806
\bibitem[Wang \& White(2012)]{Wang:2012} Wang, W., \& White, S.~D.~M.\ 2012,
  arXiv:1203.0009
\bibitem[Weil \& Hernquist(1996)]{Weil:1996} Weil, M.~L., \& Hernquist, 
  L.\ 1996, \apj, 460, 101
\bibitem[Wilson et al.(2006)]{Wilson:2006} Wilson, C. D., Harris, W. E., Longden, R., \& Scoville, N. Z, 2006, \apj, 642, 763
\bibitem[Worthey \& Ottaviani(1997)]{Worthey:1997} Worthey, G., \& Ottaviani,
  D.~L.\ 1997, \apjs, 111, 377

\end{thebibliography}
\end{document}